\documentclass[11pt,preprint]{aastex}
\usepackage{epsfig,emulateapj5,apjfonts}

\newcommand       \AU           {\,{\rm AU}}

\newcommand       \g            {\,{\rm g}}
\newcommand       \K            {\,{\rm K}}
\newcommand       \pc           {\,{\rm pc}}

\newcommand       \myr          {\,{\rm Myr}}
\newcommand       \Teff         {T_{\rm eff}}

\newcommand       \hra          {{\rm HR\,4796A}}

\newcommand       \amin         {a_{\rm min}}
\newcommand       \amax         {a_{\rm max}}
\newcommand       \rin          {r_{\rm in}}
\newcommand       \rout         {r_{\rm out}}

\newcommand \mum {\,{\rm \mu m}}
\newcommand \cm {\,{\rm cm}}

\newcommand \simali {{\sim\,}}
\newcommand       \simlt        {\lesssim}

\newcommand \Fdisk {F_{\rm disk}(\lambda)}
\newcommand \Fstar {F_\lambda^{\star}}
\newcommand \Fsca  {F_{\rm sca}(\lambda)}
\newcommand \Ftherm  {F_{\rm therm}(\lambda)}
\newcommand	  \sigmap       {\sigma_{\rm p}}
\newcommand	  \sigmar       {\sigma(r)}
\newcommand	  \rp           {r_{\rm p}}
\newcommand	  \dof          {{\rm dof}}
\newcommand	  \Ndata        {N_{\rm data}}
\newcommand	  \Npara        {N_{\rm para}}

\shorttitle{Organic Matter in the HR\,4796A Disk?}

\begin{document}

\title{
        Complex Organic Materials in the HR\,4796A Disk?
	 }
\author{M. K{\"o}hler\altaffilmark{1}, 
        I. Mann\altaffilmark{2},
    and Aigen Li\altaffilmark{1}}
\altaffiltext{1}{Department of Physics and Astronomy,
                 University of Missouri, Columbia, MO 65211;
                 {\sf koehlerme@missouri.edu, LiA@missouri.edu}}
\altaffiltext{2}{School of Science and Engineering, Kindai University, 
                 Osaka 577-8502, Japan;
                 {\sf mann@kindai.ac.jp}}  

\begin{abstract}
%
The red spectral shape of the visible to near infrared 
reflectance spectrum of the sharply-edged ring-like disk
around the young main sequence star HR\,4796A was recently
interpreted as the presence of tholin-like complex organic
materials which are seen in the atmosphere and surface 
of Titan and the surfaces of icy bodies in the solar system. 
However, we show in this {\it Letter} that porous grains
comprised of common cosmic dust species (amorphous silicate, 
amorphous carbon, and water ice) also closely reproduce 
the observed reflectance spectrum, suggesting that 
the presence of complex organic materials in the $\hra$ 
disk is still not definitive. 
\end{abstract}


\section{Introduction}
$\hra$ is a nearby 
(distance to the Earth $d\approx 67\pm 3\pc$)
young main-sequence (MS) star 
(age $\approx 8\pm 3\myr$; Stauffer et al.\ 1995) 
of spectral type A0V 
(effective temperature $\Teff\approx 9500\K$)
with a large infrared (IR) excess
which has recently aroused considerable interest.
%
Imaging observations describe dust scattering in the optical 
(Debes et al.\ 2008) and near-IR (Augereau et al.\ 1999; 
Schneider et al.\ 1999) and thermal emission at mid-IR 
(Jayawardhana et al.\ 1998; Koerner et al.\ 1998; 
Telesco et al.\ 2000; Wahhaj et al.\ 2005).
They reveal a ring-like disk with maximum 
at $\simali$70$\AU$ distance from the central star 
and $\simali$17$\AU$ width that is sharply truncated  
at the inner and the outer edge.
%
The structure of the HR\,4796A disk has important implications 
for planetesimal evolution (Kenyon et al.\ 1999). Furthermore, 
possibly existing planets may generate disk asymmetries through 
gravitational confinement or perturbation (Wyatt et al.\ 1999) 
or form rings with sharp edges (Augereau et al.\ 1999,
Klahr \& Lin 2000, Th\'ebault \& Wu 2008). 

Very recently, Debes et al.\ (2008) measured a visible 
to near-IR photometric reflectance spectrum of 
the dust ring around $\hra$. To fit the observed
spectrum (which is characterized by a steep red slope
increasing from $\lambda \approx 0.5\mum$ to 1.6$\mum$
followed by a flattening of the spectrum at
$\lambda >1.6\mum$), Debes et al.\ (2008) argued
for the presence of tholin-like organic material
in the disk around $\hra$. 
Tholin, a complex organic material, was detected as 
a major constituent of the atmosphere and surface 
of Titan and the surfaces of icy bodies in the solar system. 
The detection of tholin in the $\hra$ disk 
-- if confirmed -- would imply that these potential
basic building blocks of life may be common in 
extra-solar planetary systems as well
(see van Dishoeck 2008 for an overview 
of organic matter in space).
%
Its young age places HR\,4796A at a transitional stage 
between massive gaseous protostellar disks around 
young pre-MS T-Tauri and Herbig Ae/Be stars 
($\simali$1$\myr$) and evolved and tenuous debris disks around
MS ``Vega-type'' stars ($\simali$100$\myr$; 
see Jura et al.\ 1993, Chen \& Kamp 2004).
Detecting organic matter in this system
will provide valuable information about 
the formation and evolution of planetary systems.

The observed thermal emission of the disk, 
however, was previously reproduced with a model 
population of porous grains consisting of
the common cosmic dust species amorphous silicate,
amorphous carbon, and water ice (Li \& Lunine 2003a). 
In this {\it Letter} we question the existence
of tholin as a major dust component in the HR\,4796A disk 
and study alternative dust models to reproduce 
the observed reflectance spectrum.


\section{Model\label{calculations}}
Dust in the disk around $\hra$ scatters starlight
at visible to near-IR wavelengths and emits thermally
in the IR. The total flux of the disk $\Fdisk$ 
is the sum of the scattered light $\Fsca$ 
and the dust thermal emission $\Ftherm$. 
At $\lambda < 2.2\mum$ the dominant contribution to $\Fdisk$ 
comes from the starlight scattered by dust $\Fsca$
which is calculated from
\begin{equation}
\Fsca = \frac{\Fstar}{4 \pi d^2} 
\int\limits^{\rout}_{\rin}
\left(\frac{R_{\star}}{2r}\right)^2 
\sigma(r)\,2\pi\,r\,dr
\int\limits^{\amax}_{\amin} 
C_{\rm sca}(\lambda,a)\,\Phi(\lambda,a)\,n(a)\,da ~,
\end{equation}
where $d\approx 67\pc$ is the distance from the star to the Earth; 
$\Fstar$ is the stellar atmospheric flux 
approximated by the Kurucz model for A0V stars 
with $\Teff = 9500\K$, $\lg g=4.5$ 
and a solar metallicity (Kurucz 1979);
$R_{\star} = 1.7\,R_{\odot}$ is
the stellar radius; 
$\rin$ and $\rout$ are respectively 
the inner and outer boundaries of the disk;
$\sigmar$ is the dust surface density distribution;
$C_{\rm sca}(\lambda,a)$ is the scattering cross section of 
spherical dust of radius $a$ at wavelength $\lambda$;
$\Phi(\lambda,a)$ is the phase function approximated by 
the Henyey-Greenstein function at a scattering angle of 
$\theta = 90^{\circ}$ (see Debes et al.\ 2008)
with the asymmetry parameter $g$
calculated from Mie theory for dust of size $a$;\footnote{%
  The asymmetry parameters $g$ of the best-fit models 
  averaged over the size distributions show a gradual
  decrease from $g\approx 0.96$ at $\lambda=0.55\mum$
  to $g\approx 0.85$ at $\lambda=2.5\mum$, 
  indicating that the dust is highly forward-throwing.
  We should note that one should not compare the $g$ values
  calculated here with that of Debes et al.\ (2008),
  derived from the observed surface brightness of 
  the disk together with the Henyey-Greenstein phase function.
  The former specify the degree of scattering in the forward 
  direction ($\theta = 0^{\circ}$) of the dust
  (see Li 2008),
  while the latter approximate the mean scattering properties 
  of the dust averaged over the entire disk.
  }
$n(a)$ is the dust size distribution which is taken
to be a power-law $n(a) \propto a^{-\alpha}$ 
with a lower-cutoff $\amin$, an upper-cutoff $\amax$,
and a power-law index $\alpha$.
We take $\amax =1000\mum$ and treat $\amin$ 
and $\alpha$ as free parameters.

Following Kenyon et al.\ (1999), Klahr \& Lin (2000),
and Li \& Lunine (2003a), we approximate the dust surface density 
distribution by a Gaussian-type function
$\sigmar = \sigmap \exp[-4\ln2\{(r-\rp)/\Delta\}^2]$
where $\rp$ is the radial position where $\sigmar$ peaks,
$\Delta$ is the full width half maximum (FWHM) of the distribution,
and $\sigmap$ is the mid-plane surface density at $r=\rp$.
We take $\rp=70\AU$ and $\Delta=15\AU$ 
(see \S2.1 of Li \& Lunine 2003a).
We take $\rin=40\AU$ and $\rout=100\AU$ since there is little dust
even at $r<55\AU$ or $r>85\AU$ 
(see Fig.\,5 of Li \& Lunine 2003a).

We consider porous grains composed of common dust species 
(amorphous silicate, amorphous carbon,\footnote{%
  The other carbon dust species widely considered 
  in astrophysical modeling are graphite,
  hydrogenated amorphous carbon, 
  quenched carbonaceous composite, 
  and organic refractory. 
  Their optical properties are not qualitatively different
  from that of amorphous carbon.
  }
and water ice; see Li \& Lunine 2003a). 
The optical properties of porous dust are determined with 
Mie theory in combination with the Bruggeman effective
medium theory (Bohren \& Huffman 1983; see eqs.7--9 of
Li \& Lunine [2003b] for a detailed description).
We take the mass ratio of amorphous carbon 
to amorphous silicate to be $m_{\rm carb}/m_{\rm sil} = 0.7$
and the mass ratio of water ice to amorphous carbon
and amorphous silicate to be
$m_{\rm ice}/m_{\rm carb}+m_{\rm sil} = 0.8$,
as inferred from the cosmic abundance constraints
(see Appendix A of Li \& Lunine 2003a).
Porous dust models consisting of fluffy aggregates of
amorphous silicate, amorphous carbon and water ice
with such mixing ratios have been shown successful 
in reproducing the IR to submillimeter dust emission 
spectral energy distribution of the $\hra$ disk 
(Li \& Lunine 2003a, Sheret et al.\ 2004).

For amorphous silicate dust, we assume amorphous 
${\rm (Mg,Fe) SiO_{4}}$, an amorphous material with 
olivine-normative composition for which we take 
optical constants from J\"ager et al.\ (1994).
The optical constants of amorphous carbon and water ice
are taken from Rouleau \& Martin (1991; ``AC'' type) 
and Warren (1984), respectively. 
Crystalline silicates are not included in our model 
calculations since observations at $\simali$8--13$\mum$ 
with the mid-IR Keck LWS show no crystalline silicate 
emission features (Kessler-Silacci et al.\ 2005).
Although the optical properties of crystalline silicates 
differ from that of amorphous silicates,\footnote{%
  While crystalline silicates in astrophysical regions 
  are often found to be Mg-rich and Fe-poor 
  (e.g. see Molster \& Kemper 2005), 
  astronomical amorphous silicates appear to 
  have a similar fraction of Mg and Fe as implied 
  by the strong UV/visible absorptivity 
  required to model the circumstellar emission 
  (Jones \& Merrill 1976, Rogers et al.\ 1983) 
  and interstellar extinction (Draine \& Lee 1984).
  This suggests that amorphous silicates could
  be less effective in scattering the visible to near-IR
  starlight.
  }
the resulting reflectivity is similar. 
Moreover, the amount of crystalline silicates,
if observed in disks, is significantly smaller 
than that of the amorphous silicates. 

A major characteristic of porous dust is its porosity $P$
(i.e. the fractional volume of vacuum in a porous grain).
Li \& Lunine (2003a) have shown that the IR emission of
the HR\,4796A disk is best fit by dust with $P=0.90$.
We should note that the best-fit porosity of $P=0.90$
refers to the porous aggregates of amorphous silicate
and amorphous carbon; the porosity is reduced to $P\approx 0.73$
when ice fills in some of the vacuum under the assumption
of a complete condensation of all condensable volatile elements
as ice (see Appendix B of Li \& Lunine 2003a).
It is expected that the dust in the $\hra$ disk 
will be coated by ice since at $\simali$70$\AU$ from the star
the dust will be cooler than $\simali$110--120$\K$ 
and ice condensation will occur (see Li \& Lunine 2003a).
Assuming a complete ice condensation,
the original porosity $P$ for the porous mixture of
silicate and carbon will be reduced to
${\rm max}\left\{0,\,\left[1-2.68\,(1-P)\right]\right\}$. 
Unless stated otherwise, in the following the porosity $P$ 
refers to the fractional volume of vacuum in ice-coated 
porous dust. We will consider a range of porosities:
$P=0, 0.2, 0.5, 0.73, 0.9$
(which correspond to $P\approx 0.63, 0.70, 0.81, 0.90, 0.96$
if ice is removed).

\section{Results} \label{results}
We compare $\Fsca/\Fstar$ with the visible to near-IR
reflectance spectrum of the $\hra$ disk compiled by
Debes et al.\ (2008) either from archival images
(HST/STIS 50CCD at $\lambda$\,=\,0.585$\mum$,
HST/NICMOS F110W at $\lambda$\,=\,1.1$\mum$, 
F160W at $\lambda$\,=\,1.6$\mum$)
or from their newly obtained HST/NICMOS images
(F171M, F180M, F204M and F222M 
at $\lambda$\,=\,1.71, 1.80, 2.04, 2.22$\mum$,
respectively).  

\begin{figure}
\begin{center}
\includegraphics[width=12.8cm]{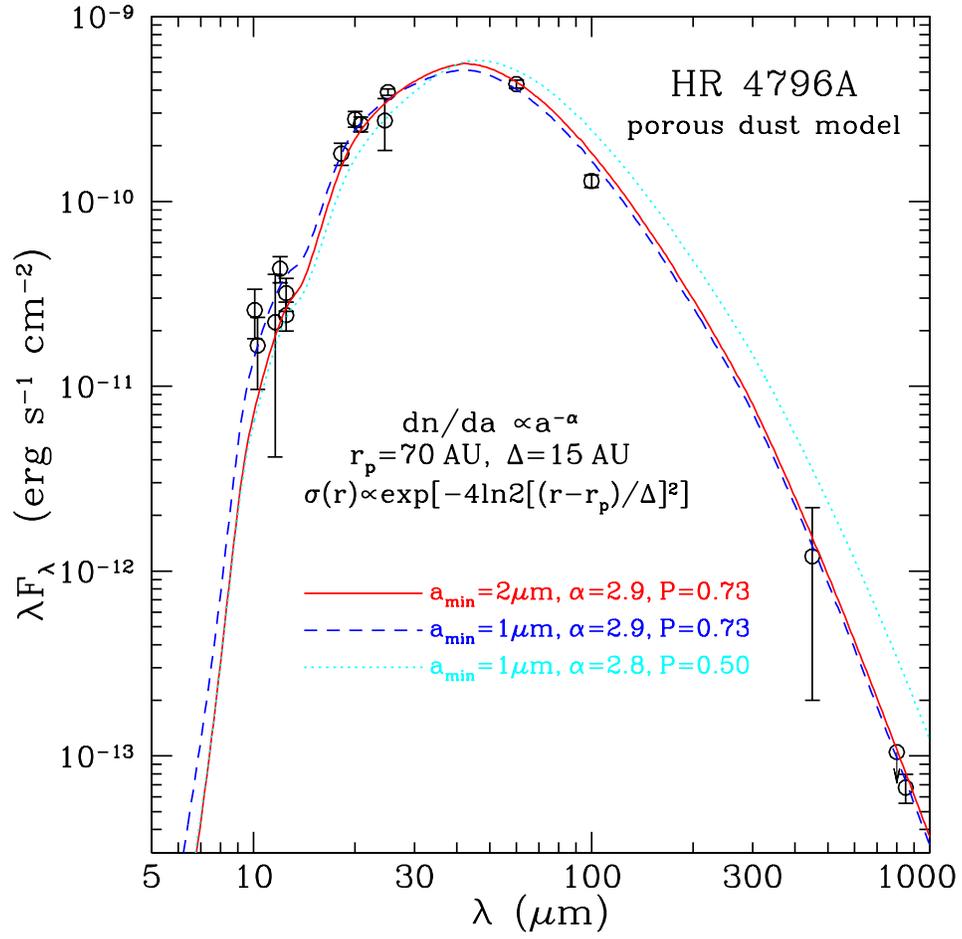}
\end{center}
\caption{\label{fig:kml_irem}
         Comparison of the observed IR emission
         of the HR\,4796A dust disk to the model 
         spectra calculated from the porous dust models 
         consisting of amorphous silicate, amorphous carbon 
         and ice. 
         }
\end{figure}

By varying $\amin$ and $\alpha$, we try to fit
the observational data of Debes et al.\ (2008)
with the porous dust model consisting of common dust species
for which the mass mixing ratios are taken from the cosmic
abundance considerations (see \S2).
We first take the porosity to be $P=0.73$ 
(which corresponds to $P=0.90$ for the porous dust
without ice coating; the model with $P=0.90$ gives 
the best fit to the observed IR emission, as shown
in Li \& Lunine 2003a).
As illustrated in Figure \ref{fig:kml_refl},
the model with $P=0.73$, $\amin\approx 2\mum$, 
and $\alpha\approx2.9$ closely reproduces   
the reflectance spectrum of $\hra$. 
In Figure \ref{fig:kml_dev} we plot the deviations of 
the model predictions from the HST photometry at each
of the seven bands. The deviations are smaller than
$\simali$7.5\% for all wavebands.
It is even more encouraging that the same model also
closely fits the IR emission of this disk
(see Fig.\,\ref{fig:kml_irem}), except its
slight deficiency at $\simali$10$\mum$ (but still
well within the observational uncertainties).
The model requires a total dust mass of
$\simali$4.27$\times10^{27}\g$
and results in a vertical visible optical depth
of $\tau_V\approx 0.048$ at $r=\rp$
and a mid-plane radial optical depth
of $\tau_V\approx 0.56$.
This justifies the optical thin approximation
employed in this work.

In contrast, the model with $P=0.73$, $\amin\approx 1\mum$, 
and $\alpha\approx2.9$ provides the best fit
to the observed IR emission (see Li \& Lunine 2003a
and Fig.\,\ref{fig:kml_irem}). However, this model
does not closely fit the observed reflectance spectrum
(see Figs.\,\ref{fig:kml_refl},\ref{fig:kml_dev}),  
although the model reflectance spectrum does exhibit 
a general trend (i.e. a steep red slope at $\simali$0.5--1.6$\mum$
and subsequent flattening off; see Fig.\,\ref{fig:kml_refl}) 
similar to that observed in the $\hra$ disk.

The best fit to the observed reflectance spectrum is given 
by the model with $P=0.50$, $\amin\approx 1\mum$, 
and $\alpha\approx2.8$ (see Fig.\,\ref{fig:kml_refl}). 
The deviations from the reflectance data are smaller than
$\simali$6.2\% for all wavebands (see Fig.\,\ref{fig:kml_dev}). 
However, the model does not fit the observed IR emission
(see Fig.\,\ref{fig:kml_irem}). 

Generally speaking, highly porous dust (with $P>0.85$)
does not fit the observed reflectance spectrum well,
while more compact dust (with $P<0.60$) is too cold
to reproduce the observed IR emission.
In Table \ref{tab:para} we list the parameters used 
in the different models: the minimal grain radius $\amin$, 
the exponent of the size distribution $\alpha$, 
the porosity $P$, and $\chi^2/\dof$, 
where $\dof\equiv \Ndata-\Npara$ 
is the ``degree of freedom'' ($\Ndata = 7$ is 
the number of data points at 
$\lambda$\,=\,0.585, 1.1, 1.6, 1.71, 1.80, 2.04, 2.22$\mum$,
$\Npara=3$ is the number of free parameters: 
$\amin$, $\alpha$ and $P$).

\section{Discussion} \label{discussion}
Debes et al.\ (2008) calculated the reflectance 
spectra of ``astronomical silicates'', water ice,
hematite Fe$_2$O$_3$ (which is found on the surface of Mars
and responsible for its redness), and
UV laser ablated olivine (which has been used to 
explain the spectral reddening of silicate-rich asteroids 
due to space weathering; Brunetto et al.\ 2007).
But the scattering spectra of these minerals and ice 
are too neutral at $\lambda$\,$\sim$\,0.5--1.6$\mum$
to match the steep red spectral slope of the reflectance 
spectrum of the $\hra$ disk.
Debes et al.\ (2008) therefore resorted to organic materials.
They found that tholin or its mixture with other dust species
(e.g. water ice or olivine) are able to reproduce
the observed red spectral slope.
This led them to suggest that
``{\it the presence of organic material is the most 
plausible explanation for the observations}'',
with a cautionary note that ``... {\it longer wavelength 
scattered light observations will further constrain 
the (tholin-based) grain models,
particularly around 3.8--4$\mum$, where a large absorption 
feature is seen for different grain sizes of tholins. 
This would help to directly confirm whether Titan tholins 
are an adequate proxy for the material in orbit around 
HR\,4796A.}''

\begin{figure}
\begin{center}
\includegraphics[width=12.8cm]{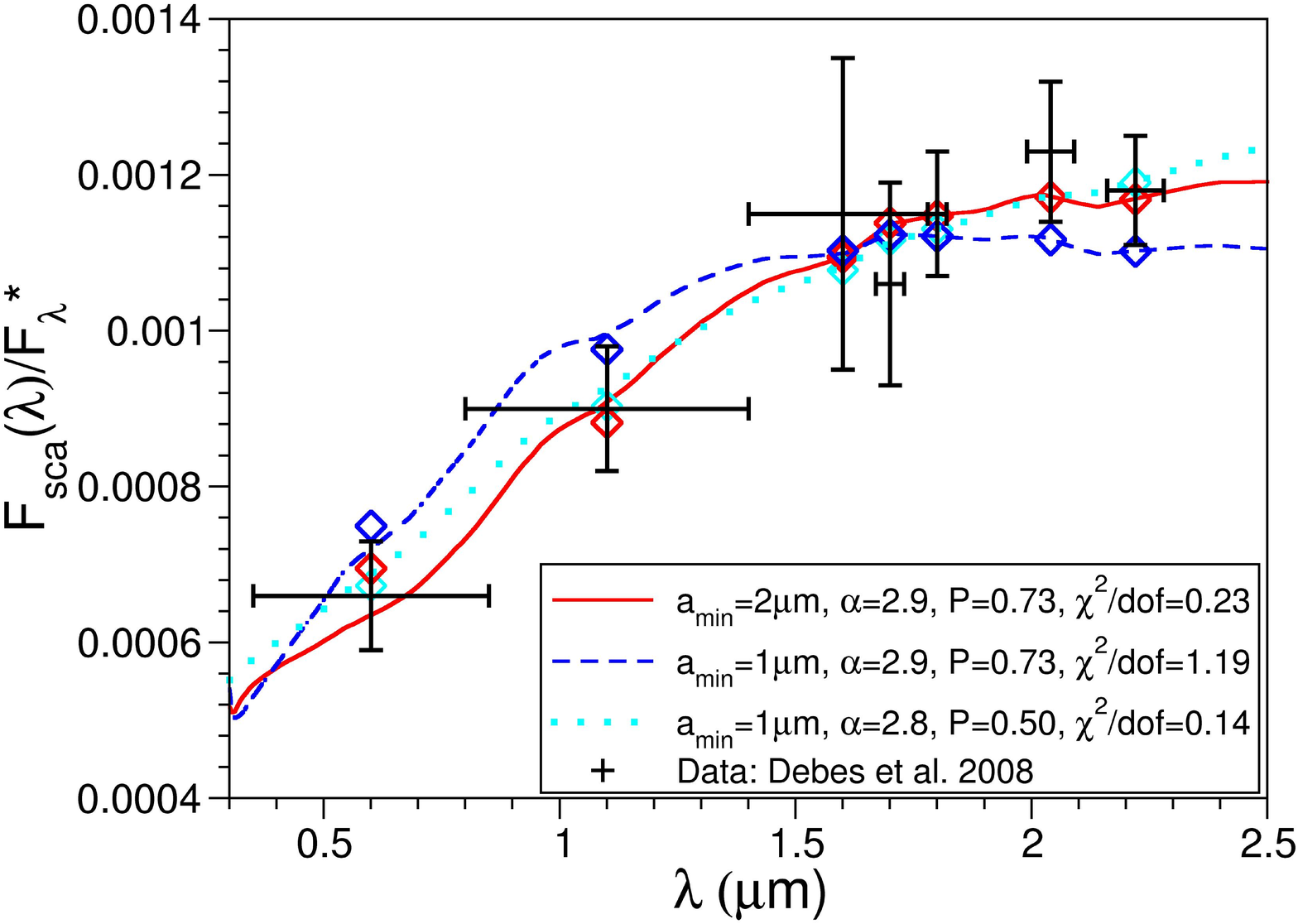}
\end{center}
\caption{\label{fig:kml_refl}
         Comparison of the model scattered light 
         spectra with the observed visible to near-IR 
         reflectance spectrum of the $\hra$ disk
         of Debes et al.\ (2008; crosses).
         Diamonds show the model spectra convolved
         with the HST STIS and NICMOS filters. 
         The model with $P=0.73$, $\amin=2\mum$
         and $\alpha=2.9$ (red solid line) is preferred 
         since it fits both the observed reflectance spectrum
         and the observed IR emission (see Fig.\,\ref{fig:kml_irem}).
         While the model with $P=0.50$, $\amin=1\mum$
         and $\alpha=2.8$ (cyan dotted line) fits the
         reflectance spectrum better than any other models,
         it does not fit the observed IR emission 
         (see Fig.\,\ref{fig:kml_irem}).       
         In contrast, the model with $P=0.73$, $\amin=1\mum$
         and $\alpha=2.9$ (blue dashed line) provides an
         excellent fit to the observed IR emission,
         its fit to the observed reflectance spectrum
         is not as good as the other two models shown
         in this figure.
        }
\end{figure}

\begin{figure}
\begin{center}
\includegraphics[width=12.8cm]{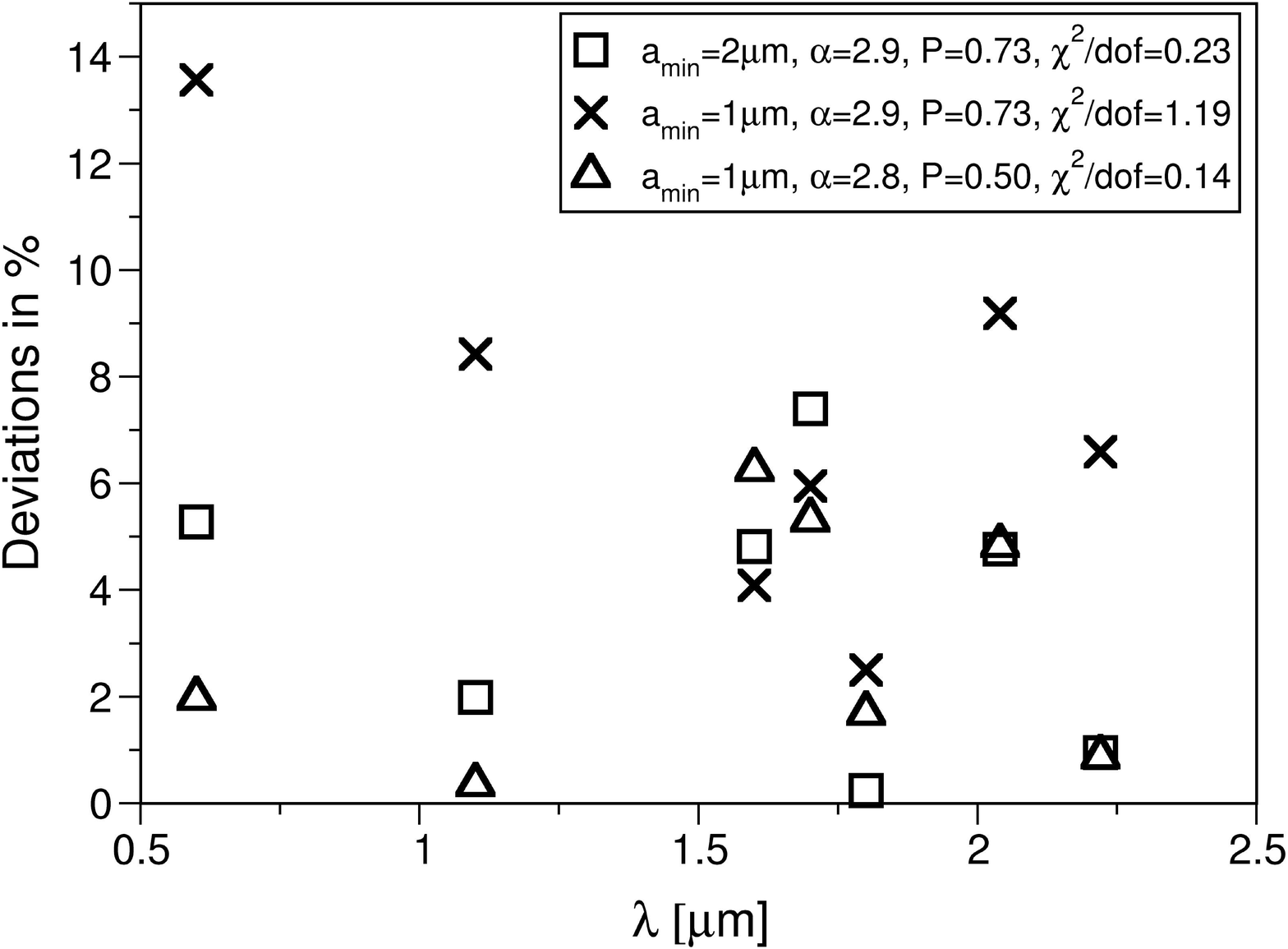}
\end{center}
\caption{\label{fig:kml_dev}
         Deviations 
   $\left|\left[\Fsca/F_\lambda^{\star}\right]_{\rm mod}/
   \left[\Fsca/F_\lambda^{\star}\right]_{\rm obs} - 1\right|$
         of the model results
         (integrated over the instrument filters)
         from the HST STIS and NICMOS photometry 
         at each of the seven bands. 
         The deviations are within 
         $\simali$7.5\% at all wavebands
         for the preferred model
         (with $P=0.73$, $\amin=2\mum$ and $\alpha=2.9$)
         which closely fits both the reflectance spectrum
         (see Fig.\,\ref{fig:kml_refl})
         and the IR emission of the $\hra$ disk
         (see Fig.\,\ref{fig:kml_irem}).
         }
\end{figure}

However, as shown in Figure \ref{fig:kml_refl},
simple porous dust models consisting of dust species
(amorphous silicate, amorphous carbon, water ice) 
which are commonly considered to dominate in 
the interstellar medium (ISM), envelopes around
evolved stars, and dust disks around young stars
closely reproduce the observed reflectance spectrum
of the $\hra$ disk. 
Our model provides at least a viable alternative 
to the tholin-based models of Debes et al.\ (2008).
While the tholin organic dust model predicts a strong 
feature around 3.8--4$\mum$ characteristic of tholin 
(Debes et al.\ 2008), the porous dust model presented
here predicts a strong band at $\simali$3.1$\mum$, 
attributed to the O--H stretching mode of water ice.
 
The tholin organics of which the optical constants 
were adopted by Debes et al.\ (2008) were made from 
DC discharge of 90\% N$_2$ and 10\% CH$_4$ gas mixture
(Khare et al.\ 1984). They are extremely N-rich and
optically very different from amorphous carbon.
The dust in the HR\,4796A disk should be continuously
replenished. This is indicated by the low size cutoff 
$\amin$ (of a few micrometers) of the dust required to 
reproduce the reflectance spectrum 
(see \S2 and Table \ref{tab:para})
combined with considerations of dust lifetimes 
based on the radiation pressure and Poynting-Robertson effects 
(see Fig.\,9 of Li \& Lunine 2003a).
The replenishing source would likely arise from 
collisional cascades of larger bodies like planetesimals,
asteroid-like and comet-like bodies. 
We suggest, with interstellar dust as the building blocks of 
the parent bodies, it is more reasonable to assume
that the dust in the HR\,4796A disk is composed of amorphous silicate,
amorphous carbon, and water ice.\footnote{%
  The tholin model may still be viable if the dust in 
  the $\hra$ disk originates from the surface layers 
  of planetesimals, possibly through excavating collisions
  (J.H. Debes, private coomunication).
  In this scenario, the entire planetesimal need
  not be composed of tholins; they might reside only on 
  the surface where they are created and then be released. 
  Due to significant processing and 
  possible alteration through planetesimal formation,  
  the surface compositions of planetesimals
  in the $\hra$ disk may not resemble the ISM composition.
  Based on what we know from our own Solar System, 
  methane ice and water ice may reside on the surfaces of large
  planetesimals. These ices are exposed to the stellar UV flux 
  and may ultimately produce tholin-like organic residues.
  }

The model which best fits both the observed reflectance 
spectrum and the IR emission requires highly porous dust
(with $P=0.73$ which corresponds to $P=0.90$ if ice is
sublimated; see \S2). While it is natural to recognize that 
cold conglomeration of dust grains in molecular clouds can 
lead to highly porous dust structures,\footnote{%
  A porosity in the range of $0.80\simlt P\simlt 0.97$
  is expected for dust aggregates formed through coagulation
  as shown both theoretically 
  (Cameron \& Schneck 1965; Wada et al.\ 2008) 
  and experimentally (Blum et al.\ 2006).
  }
at a first glance, it is harder to accept that comparatively 
more violent collisions between larger bodies would result in 
such a morphology in the resulting debris.

To address this concern, we take the interplanetary dust 
particles (IDPs) as an analog for the dust in debris disks. 
The anhydrous chondritic IDPs collected in the stratosphere 
possibly of cometary origin show a highly porous structure 
(Brownlee 1987). Love et al.\ (1994) have measured the densities
of $\simali$150 unmelted chondritic IDPs with diameters 
of $\simali$5--15$\mum$, using grain masses determined 
from an absolute X-ray analysis technique with a transmission
electron microscope and grain volumes determined from
scanning electron microscope imaging.
They found that these particles have an average density 
of $\simali$2.0$\g\cm^{-3}$, corresponding to a moderate 
porosity of $\simali$0.4.
More recently, Joswiak  et al.\ (2007) identified 12 porous 
cometary IDPs (based on their atmospheric entry velocities)
with an average density of $\simali$1.0$\g\cm^{-3}$,
corresponding to $P\approx 0.7$.
Much higher porosities ($>$0.9) have been reported
for some very fluffy IDPs (e.g. MacKinnon et al.\ 1987,
Rietmeijer 1993), despite that highly porous IDPs are 
probably too fragile to survive atmospheric entry heating.
Low densities are also derived for different groups of
meteoroids\footnote{%
  By definition, meteoroids are small bodies in the mass range 
  of $\simali$$10^{-4}$--$10^8\g$, 
  which orbit the Sun in interplanetary space. 
  The atmospheric trajectories of meteors, i.e. the brightness 
  generated by meteoroids passing through atmosphere, contain
  information about meteoroid orbits and densities. 
  Meteoroids are accordingly classified into groups
  with different orbits, structure and composition. 
  The densities given above are valid for between 
  $\simali$50\% and 73\% of cometary meteor observations, 
  depending on the observation method. 
  The same studies show densities 
  of $\simali$2$\g\cm^{-3}$ for the remaining cometary meteoroids, 
  as well as for a significant part of the meteoroids ascribed to
  asteroids, which is also moderately porous
  (see Mann 2008).
  }
ascribed to comets: the densities are 
$\simali$1.0, 0.75, and 0.27$\g\cm^{-3}$, 
respectively (Ceplecha et al.\ 1998),
corresponding to a porosity of $\simali$0.7, 0.8, and $>$0.9.


It is therefore reasonable to assume that the dust in
the $\hra$ disk has a high porosity (at least for those
originated from cometary bodies). We are not sure about
the relative contributions to the $\hra$ disk from 
asteroid collisions and cometary activity. 
Note that this is a long-standing problem even
for our own solar system (e.g. see Lisse 2002). 

To conclude, we argue that the presence of tholin-like 
complex organic materials in the $\hra$ disk is still
not conclusive since the observed red spectral shape
of the disk can be closely reproduced by models of
porous dust comprised of common cosmic dust species.
A more thorough study of the scattered light
over a range of scattering angles would further
constrain the optical properties of the dust.

\acknowledgments
We thank J.H. Debes and the anonymous referee 
for their very helpful comments.
MK and AL are supported in part
by NASA/HST Theory Programs
and NSF grant AST 07-07866.
AL is supported by the NSFC
Outstanding Overseas Young Scholarship.


\begin{table}
\centering
\begin{minipage}{140mm}
\caption{Parameters for porous dust models consisting 
           of common dust species (amorphous silicate, 
           amorphous carbon, water ice).
           Model 5 is preferred since it closely
           reproduces both the observed reflectance
           spectrum (see Fig.\,\ref{fig:kml_refl})
           and the observed IR emission
           (see Fig.\,\ref{fig:kml_irem}). 
           Although model 3 fits the observed reflectance
           spectrum better than any other models 
           (see Fig.\,\ref{fig:kml_refl}), it does not
           fit the observed IR emission
           (see Fig.\,\ref{fig:kml_irem}). 
           \label{tab:para}
           }
\begin{center}
\begin{tabular}{ccccc}
\hline
Dust  & $\amin$  & $\alpha$ & Porosity 
      & $\chi^2/{\rm dof}$ \\
Model & ($\mu$m) &          &  (\%)  & \\
\hline
1     &  1 & 2.8   &   0    &  0.47  \\
2     &  1 & 2.8   &   20   &  0.33  \\
3     &  1 & 2.8   &   50   &  0.14  \\
4     &  1 & 2.9   &   73   &  1.19  \\
{\bf 5}     &  {\bf 2} & {\bf 2.9}   & {\bf 73}   &  {\bf 0.23}  \\
6     &  4 & 3.2   &   90   &  1.74  \\
\hline
\end{tabular}
\end{center}
\end{minipage}
\end{table}

\end{document}